\documentstyle[11pt,newpasp,twoside,epsf]{article}
\def\edcomment#1{\iffalse\marginpar{\raggedright\sl#1\/}\else\relax\fi}
\reversemarginpar

\begin{document}
\title{Radio and Infrared Properties of Dust-Enshrouded Wolf-Rayet Stars}
 \author{J. D. Monnier \& L. J. Greenhill}
\affil{Harvard-Smithsonian Center for Astrophysics, 60 Garden St, Cambridge, 
MA 02138, USA}
\author{P. G. Tuthill}
\affil{Chatterton Astronomy Dept., School of Physics, University of Sydney,
NSW 2006, Australia}
\author{W. C. Danchi}
\affil{NASA Goddard Space Flight Center, Infrared Astrophysics, Code 685,
Greenbelt, MD 20771, USA}

\begin{abstract}
This paper discusses our ongoing efforts to characterize
dust-enshrouded Wolf-Rayet (WR) stars in the radio and infrared. 
We have used the
Very Large Array to measure the broadband radio
spectrum of WR stars in suspected binary systems and 
discovered non-thermal emission, which is usually attributed to colliding winds.
In addition, infrared imaging using aperture masking
interferometry on the Keck-I
telescope has resolved the dust shells around a number of WR stars
with K-magnitudes brighter than $\sim$6.  Although this 
admittedly small study 
suffers from selection bias,
we note that all the dust-enshrouded WR stars with radio
detections show evidence for colliding
winds, supporting the theory that wind compression in a binary system
is {\em necessary} for efficient dust production.  A consequence of this
hypothesis is that virtually all WC8-10 stars must be in binaries, since
most are dusty.  Single-star and binary stellar evolution models will
have to be modified to accommodate this observational result if confirmed.
\end{abstract}

\section{Introduction}
The relationship of binarity to 
dust production for Wolf-Rayet (WR) stars is
controversial.  Most will agree that the physical conditions around
single WRs are unfavorable for dust production 
(e.g., Cherchneff \& Tielens 1995; Le Teuff et al. 2001) due to the intense 
radiation field, low hydrogen content, and rapid density decrease in the
fast expanding winds, although some mechanisms have been
proposed (Zubko 1998).  The realization that 
WR 140 was an eccentric binary system naturally explained the
observed non-thermal radio emission and episodic dust production in terms of
colliding winds, with high gas compression at periastron catalyzing dust
creation (Moffat et al. 1987; Williams et al. 1990; Usov 1991).

But are colliding winds {\em necessary} for dust production around WR
stars?  Dust-enshrouded WR stars have been classified as either
``variable'' or ``persistent'' dust-producers, based on the
variability of IR flux (Williams \& van der Hucht 1992).  While the
variable dust producers can be easily understood in terms of 
WR~140-style colliding wind systems, the binary nature of the
persistent dust producers has been unknown until more
recently. Diffraction-limited infrared imaging with the Keck-I telescope
revealed that spiral-shaped dust shells, ``pinwheel'' nebulae, surround
two of the dustiest WR stars (WR 104, WR 98a),
most likely created
through the colliding winds of a WR+OB binary system in a close orbit 
(Tuthill, Monnier, \& Danchi 1999; Monnier, Tuthill, \& Danchi 1999).

These observations have led to a more unified picture of the dusty
Wolf-Rayet stars in terms of interacting wind (binary) systems.  On
the one hand, WR~140 and other episodic emitters consist of eccentric
systems with $\ga$10~year orbits, while WR~104 and WR~98a 
have more circular (circularized?) orbits
with periods $\sim$1~year.  But are all dusty sources somewhere on
this continuum of binary orbits, or could some of them be single
stars?

In this paper, we give a progress report including preliminary results
of our work to detect non-thermal emission from pinwheel nebulae and our search
for more pinwheel nebulae in the infrared.

\section{New Radio Observations}
We used the Very Large Array (VLA) to measure the broadband spectra of
three dust-enshrouded Wolf-Rayet systems, WR~104, WR~98a, and WR~112,
at wavelengths of 1.3, 2, 3.6, 6, \& 21\,cm.  At the time of our
observations, WR~112 was the only dust-enshrouded WR star to have a
radio detection (Leitherer et al. 1997; Chapman et al. 1999), and our
programme was designed to improve upon the sensitivity limits of
previous surveys by factors of 3 to 10, with a goal of
$\la$100\,$\mu$Jy point-source sensitivity.  In addition, another
12~sources have been observed with the Australian Telescope Compact
Array with somewhat less sensitivity.

\begin{figure}[tb]
\begin{center}
\plotfiddle{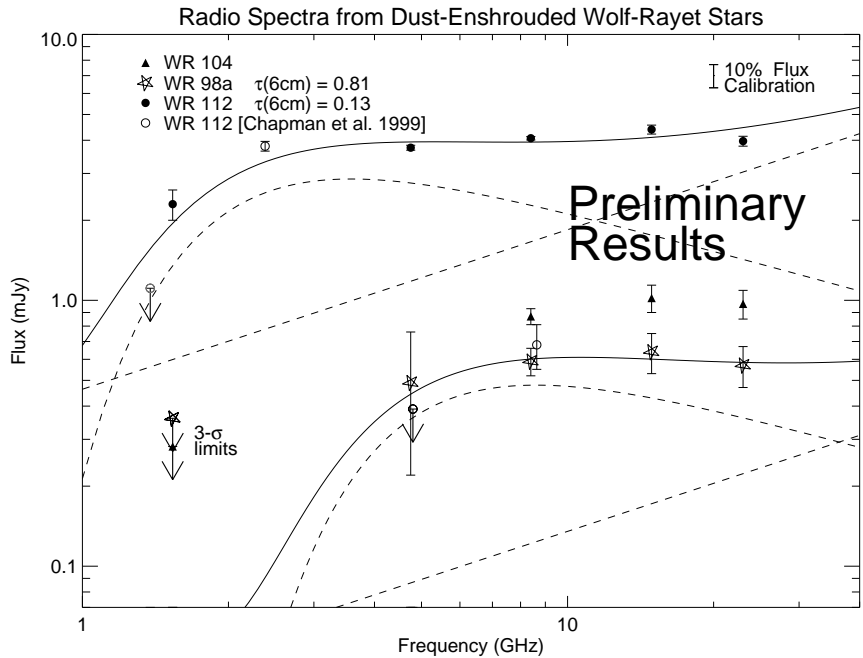}{3.2in}{90}{50}{50}{190}{-30}
\caption{Preliminary radio spectra of dust-enshrouded Wolf-Rayet stars,
including pinwheel nebulae WR 104 and WR 98a.  The solid lines 
are fits to a simple
two-component model (each component is separately plotted in a dashed linestyle),
described in text.
\label{fig:radio}}
\end{center}
\end{figure}

Preliminary results of observations from 1999 September and 2000
February are shown in Figure~1.  We have clearly detected WR~104 and
WR~98a for the first time, at the higher frequencies.  Together the
two observing sessions cover 5 wavelengths for each star. Unfortunately, we could
not study each star at all wavelengths during each epoch, but repeated
observations at certain wavelengths confirm that the fluxes did not
change dramatically between observations.  Full details will be
included in a future paper.

We draw the 
important conclusion that these sources
all show evidence for non-thermal emission.  The straight
dashed lines in Figure~1 show the expected power law slope
($\alpha\sim0.6$) for a pure (thermal) wind source, and our data does not
show the expected increase with frequency (e.g., Leitherer \& Robert 1991).  
In addition, WR~104 and WR~98a both show a dramatic drop in flux at
21~cm where the non-thermal emission should be bright,
likely caused by free-free absorption.
The radio photosphere is
expected to be well outside the non-thermal emission region at these
wavelengths, and such absorption is observed at long wavelengths
for WR~140 (White \& Becker 1995), although a detailed undertanding is
geometry-dependent.  

We fitted
a simple two-component model to the data, 
consisting of thermal
wind (spectral index $\alpha=0.6$) and non-thermal 
($\alpha=-0.5$) components, where we account for free-free
opacity, $\tau_\nu\propto\nu^{-2.1}$ (as Chapman et al. 1999).
The results of fits to the WR~112 and WR~98a data have been included
in Figure~1, and the 6\,cm optical depths are reported in the legend.
Based on the long wavelength turnover, the optical depth of the wind
appears higher for both WR~98a and WR~104, than for 
WR~112 at this epoch.
This could mean that the binary separation of WR~112 is larger than
for WR~98a and WR~104, or that we are
viewing the system through a ``hole'' in the wind.  For instance, if
the OB-type companion was in front of WR~112, then we would view the
shock collision zone through the OB-wind rather than the denser WR
wind; this kind of effect has also been seen in WR~140 (White \& Becker
1995).

The variability of the WR~112 radio flux reported in Chapman et
al. (1999), and further supported by our new data, indicates an
eccentric orbit and/or an ``edge-on'' viewing angle of
the underlying binary.  We are currently
monitoring the broadband radio spectrum of WR~112 every two months,
and hope to constrain the period and geometry by combining this
information with IR images (to be discussed in the next section) and
possible VLBA images of the wind-collision shock.  
We note that, as of 2000 June, the radio flux of WR~112 is
about a factor 2 (more at longer wavelengths) lower than seen in 2000
February.

\section{Infrared Observations}
Diffraction-limited imaging in the infrared
(resolution$\la$0\farcs050) was performed using aperture masking
techniques on the Keck-I telescope.  Observing details can be found
elsewhere in this volume (Tuthill et al. 2001), and in the literature
(Tuthill et al. 2000; Monnier 1999).  After the initial discoveries of
pinwheel nebulae around WR~104 (Tuthill et al. 1999) and WR~98a
(Monnier et al. 1999), we enlarged our sample to include most WR
stars with K-magnitude brighter than K=6.  This sample consists mostly
of late-type WC stars (WC8-10), categorized as persistent dust
emitters by Williams \& van der Hucht (1992). 

\begin{figure}[htb]
\begin{center}
\plotfiddle{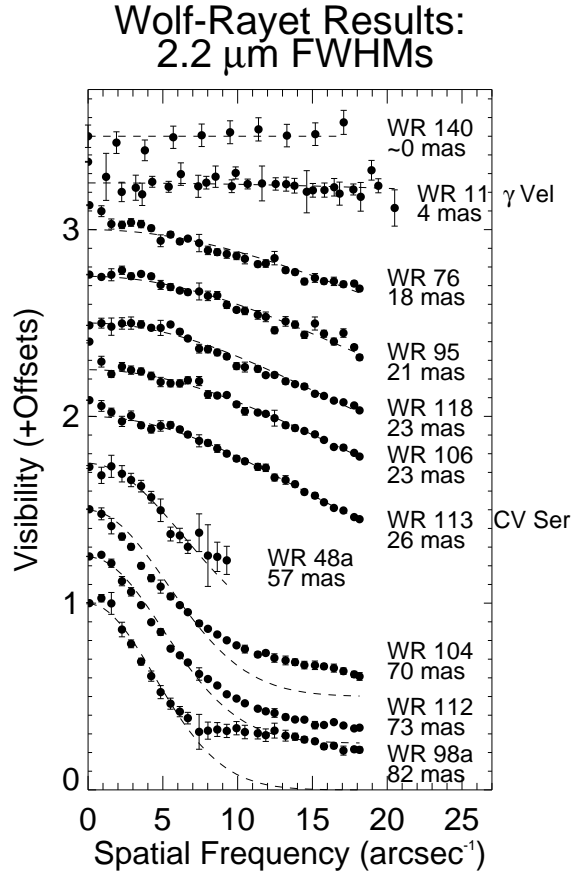}{4.0in}{0}{50}{50}{-180}{-50}
\caption{Preliminary visibility data (azimuthally averaged)
of infrared-bright Wolf-Rayet stars.  Characteristic sizes of the largest
scale emission are reported, assuming Gaussian brightness distributions.
\label{fig:vis}}
\end{center}
\end{figure}

All of the dusty sources are at least partially resolved with
sizes between 20 and 80~mas, assuming Gaussian brightness
distributions and azimuthal symmetry (see Figure~2).
A few sources (WR 140 and WR 11) do not
show evidence for dust in their spectrum and our observations confirm
that these sources are not extended.

Four sources are extended enough to allow meaningful image 
reconstructions: WR~104, WR~98a, WR~112, and WR~48a. 
While WR~104 and WR~98a have already been subjects of recently published
papers based on aperture masking, the
structure of the dust shell of WR 112 was first resolved using
lunar occultations (Ragland \& Richichi 1999). In a future paper
(Monnier et al. 2001), we will present our multi-epoch images of WR 112
showing a very disorganized and asymmetric dust shell changing significantly
with time, rather unlike the pinwheel nebulae of WR~104 and WR~98a.
WR~48a appears to be quite extended, but high resolution
imaging is hampered by telescope vignetting and poor atmospheric stability
when observing this low declination source (maximum elevation $\sim$8\deg
at Keck!).

Since we have surveyed all known IR-bright Wolf-Rayet stars and because
the emission at 2.2\micron~is dominated by thermal emission, we do not
expect to discover any more Wolf-Rayets systems with large-scale dust
emission that can be mapped using the Keck-I telescope
at this wavelength.  This is
simply because the expected angular size is proportional to the received flux,
for a given surface brightness (i.e., dust temperature).  However,
there are likely variable sources which produce dust periodically
(e.g., WR 140 in 2001 April), and these could be observed at
IR-maximum once identified.

\section{Conclusions}

\begin{itemize}
\item{The reddest (most obscured) WR stars (WR 104, WR 98a, WR 112) are in
binaries.}
\item{Non-thermal (i.e., flat spectrum) radio emission from colliding
winds in these systems is detectable, although weak, confirming the
binary hypothesis, and offering another way to establish the binary
nature of systems which are too distant to resolve with direct imaging.}
\item{WR 112 shows time variable infrared and radio emission, indicating an 
eccentric orbit for the underlying binary.  IR imaging does not reveal a 
pinwheel nebula, but 
rather an extended and dynamically changing asymmetric envelope.}
\end{itemize}

\section{Future Work}

\begin{itemize}
\item{Conduct coordinated 
VLA/VLBA monitoring and
IR imaging at Keck to answer the questions: why is this
dust shell so complex? what are the physical parameters of the underlying binary?}
\item{Determine if other obscured WR systems are in binaries.  
Methods to detect binarity include:
monitoring IR light curves, measuring radio broadband spectra, 
visible and IR spectroscopy, detection of 
X-ray signature of colliding winds (?), precise radial velocity work (?)}
\item{Since the vast majority of WC8-10 stars produce dust
(Williams et al. 1987), then 
establishing that ``all'' dusty WRs are in binaries would futher imply that 
WC8-10 stars are not produced through single star evolution.  How could 
binary evolution affect the WR precursors to produce the 
photospheric structure of the late WCs?  Perhaps Roche lobe 
overflow during a previous red supergiant phase (e.g., Monnier et al. 1999)
could play a role here. }
\item{With improved measurements of wind and binary parameters, 
accurate modelling of the high quality imagery of spiral structure
could yield new distance estimates and probe mass-loss and 
colliding wind physics.}
\item{Pursue IR interferometric observations
to significantly resolve and image dust shells
around other WR stars (possible with CHARA, IOTA). A number of promising
candidates have been identified herein.}
\end{itemize}


\end{document}